\documentclass[sigconf]{acmart}
\usepackage{graphicx}
\usepackage{subcaption}
\usepackage[ruled]{algorithm2e}
\usepackage{diagbox}
\usepackage{booktabs}
\usepackage{arydshln}
\usepackage{multirow}

\AtBeginDocument{%
  \providecommand\BibTeX{{%
    \normalfont B\kern-0.5em{\scshape i\kern-0.25em b}\kern-0.8em\TeX}}}

\copyrightyear{2025}
\acmYear{2025}
\setcopyright{acmlicensed}\acmConference[SIGIR '25]{Proceedings of the 48th International ACM SIGIR Conference on Research and Development in Information Retrieval}{July 13--18, 2025}{Padua, Italy}
\acmBooktitle{Proceedings of the 48th International ACM SIGIR Conference on Research and Development in Information Retrieval (SIGIR '25), July 13--18, 2025, Padua, Italy}
\acmDOI{10.1145/3726302.3730044}
\acmISBN{979-8-4007-1592-1/2025/07}

\begin{document}

\title{Multi-Modal Multi-Behavior Sequential Recommendation with Conditional Diffusion-Based Feature Denoising}

\author{Xiaoxi Cui}
\affiliation{
    \institution{Takway.AI}
    \city{Beijing}
    \country{China}
}
\email{cxxneu@163.com}
\authornote{These authors contributed equally to this work.}

\author{Weihai Lu}
\affiliation{
    \institution{Peking University}
    \city{Beijing}
    \country{China}
}
\email{luweihai@pku.edu.cn}
\authornotemark[1]
\authornote{Corresponding author.}

\author{Yu Tong}
\affiliation{%
  \institution{Wuhan University}
  \state{Wuhan}
  \country{China}
}
\email{yutchina02@gmail.com}

\author{Yiheng	Li}
\affiliation{%
  \institution{Shanghai University of International Business}
  \state{Shanghai}
  \country{China}
}
\email{23349096@suibe.edu.cn}

\author{Zhejun Zhao}
\affiliation{%
  \institution{Microsoft}
  \state{Beijing}
  \country{China}
}
\email{anjou1997@gmail.com}

\begin{abstract}
The sequential recommendation system utilizes historical user interactions to predict preferences. Effectively integrating diverse user behavior patterns with rich multimodal information of items to enhance the accuracy of sequential recommendations is an emerging and challenging research direction. This paper focuses on the problem of multi-modal multi-behavior sequential recommendation, aiming to address the following challenges: (1) the lack of effective characterization of modal preferences across different behaviors, as user attention to different item modalities varies depending on the behavior; (2) the difficulty of effectively mitigating implicit noise in user behavior, such as unintended actions like accidental clicks; (3) the inability to handle modality noise in multi-modal representations, which further impacts the accurate modeling of user preferences. To tackle these issues, we propose a novel \textbf{M}ulti-\textbf{M}odal \textbf{M}ulti-\textbf{B}ehavior \textbf{S}equential \textbf{R}ecommendation model (M$^3$BSR). This model first removes noise in multi-modal representations using a Conditional Diffusion Modality Denoising Layer. Subsequently, it utilizes deep behavioral information to guide the denoising of shallow behavioral data, thereby alleviating the impact of noise in implicit feedback through Conditional Diffusion Behavior Denoising. Finally, by introducing a Multi-Expert Interest Extraction Layer, M$^3$BSR explicitly models the common and specific interests across behaviors and modalities to enhance recommendation performance. Experimental results indicate that M$^3$BSR significantly outperforms existing state-of-the-art methods on benchmark datasets.
\end{abstract}



\maketitle

\section{Introduction}
With the explosive growth of internet information, recommendation systems have become a crucial bridge connecting users to vast amounts of data. Sequential recommendation, which leverages users' historical interactions to predict preferences, has emerged as a significant research direction. Traditional sequential recommendation methods primarily rely on modeling single user behaviors and item features. However, in real-world scenarios, user behaviors are diverse\cite{jin2020multi,xu2023multi,wu2022multi}, such as browsing, clicking, favoriting, and purchasing, while items are rich in multi-modal information\cite{yu2023multi,tao2020mgat}, including visual, textual, and auditory features. Effectively integrating users' diverse behavioral patterns and items' rich multi-modal information to more accurately capture user interests has become a highly promising research direction in the field of sequential recommendation\cite{chen2023survey,wang2023missrec}.
\begin{figure}[htbp]
\centering

\includegraphics[width=0.45\textwidth]{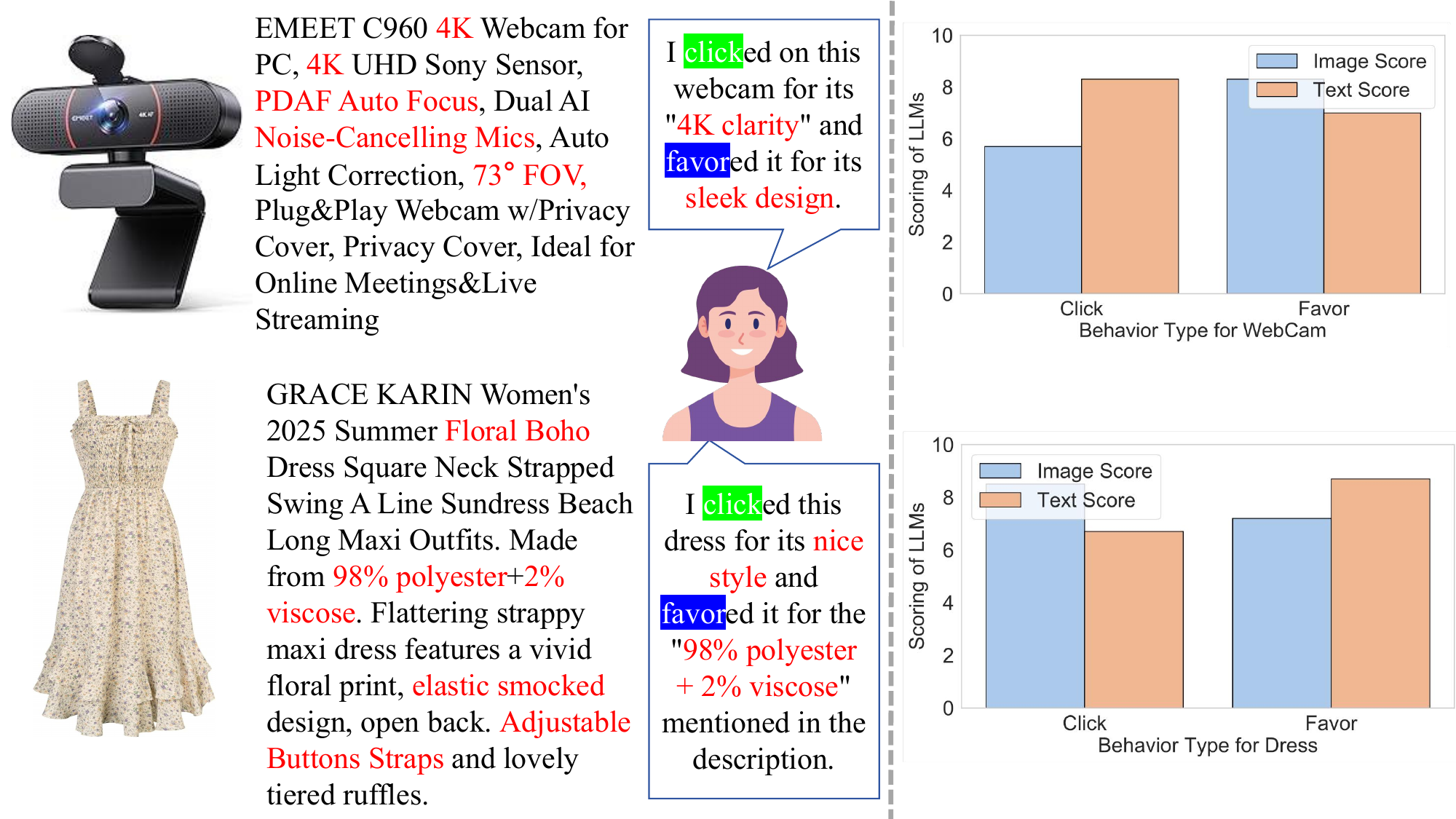}
    
    \caption{The importance of images and text in influencing user behavior varies depending on the type of behavior.  On the left, a real-world example is presented, demonstrating that user actions such as "click" and "favor" exhibit certain differences in relation to the modality of the item.  On the right, the evaluation results from multiple multimodal large language models (GPT-4o, Claude-3.5-Sonnet, and Gemini 1.5 Pro Exp-1206) are displayed, showing the statistical scores for the attractiveness of the item's image and text.  Higher scores indicate greater content attractiveness.  From this analysis, it can be observed that for the item "Dress," the image typically attracts users to "click," while the final decision to "favor" relies equally on textual information such as material and composition.}
    \label{fig:scores_comparison}
    
\end{figure}

In recent years, significant progress has been made in both multi-behavior sequential recommendation and multi-modal sequential recommendation. Multi-behavior sequential recommendation aims to leverage users' interaction information across different behaviors to more comprehensively characterize user interests~\cite{gao2019neural, yang2022multi, yang2022multi, li2024mhhcr, he2022novel}. Multi-modal sequential recommendation focuses on integrating items' multi-modal features to enhance the quality of user interest representation and recommendation effectiveness~\cite{wang2023missrec, zhang2023beyond, liang2023mmmlp}. Despite these advancements, effectively combining multi-behavior and multi-modal information for sequential recommendation remains challenging. Specifically, existing methods still face the following critical issues:

\begin{enumerate}
\item \textbf{Lack of effective characterization of modal preferences across different behaviors}: Users exhibit varying preferences for different modalities depending on their behaviors. As illustrated in Figure \ref{fig:scores_comparison}, in click behavior, visual elements might be more attention-grabbing, whereas in save behaviors, textual details and product specifications may receive greater focus. However, the lack of comprehensive research on multi-behavior multi-modal sequential recommendation has resulted in ineffective methods for characterizing users' modal preferences across behaviors.

\item \textbf{Difficulty in effectively mitigating implicit noise in user behaviors}: Implicit user behavior data often contains noise from accidental clicks or impulsive actions\cite{wang2021denoising,liu2024denoising,xin2023improving,mo2024min,mo2025one}, which can distort the modeling of true user interests and lead to irrelevant recommendations. When a user accidentally clicks on an unwanted item, it introduces misleading noise that undermines the model's understanding of preferences. Additionally, noisy behaviors complicate the characterization of modal preferences, as they may not accurately reflect users' genuine interests. In multi-behavior and multi-modal contexts, effectively identifying and mitigating the impact of such noise on modal preference modeling is a critical challenge.

\item \textbf{Inability to handle noise in multi-modal representations}: Noise in multi-modal representations negatively impacts the modeling of user preferences. Multi-modal features, such as image and text embeddings, often contain modality-specific noise unrelated to user preferences, as observed in prior work~\cite{yu2023multi}. This noise can propagate through the recommendation pipeline, further complicating the effective integration of multi-modal and multi-behavior information.

\end{enumerate}

To address these challenges, we propose a novel \textbf{M}ulti-\textbf{M}odal \textbf{M}ulti-\textbf{B}ehavior \textbf{S}equential \textbf{R}ecommendation (M$^3$BSR) model, which aims to more precisely model users' modal preferences across different behaviors while effectively mitigating noise in behavior data and multi-modal representations. Inspired by diffusion models, M$^3$BSR first employs a conditional diffusion model to remove noise from multi-modal representations (e.g., irrelevant details and noise from pre-trained models) under the guidance of user interests, thereby obtaining more accurate multi-modal preferences. Additionally, inspired by the observation that deeper user behaviors (e.g., "favorite") represent more accurate behavioral preferences~\cite{gao2019neural, guo2019buying, yang2022multi}, M$^3$BSR uses deeper behavior information as a condition to guide the diffusion model in denoising shallow behavior information (e.g., "click"), thereby improving the signal-to-noise ratio of user behavior representations. Finally, the Multi-Expert Interest Extraction Layer comprehensively models common and specific interests across behaviors and modalities through shared and dedicated expert networks, capturing the synergistic relationships between behaviors and modalities to enhance recommendation performance.

Our contributions are as follows:
\begin{itemize}
    \item We propose a novel Multi-Modal Multi-Behavior sequential recommendation (M$^3$BSR) model to improve the accuracy of sequential recommendation. To the best of our knowledge, this is the first work to explicitly model modal preferences across different behaviors.
    \item We design a Conditional Diffusion Modal Denoising Layer, a Conditional Diffusion Behavior Denoising Layer, and a Multi-Expert Interest Extraction Layer, which effectively remove noise from both modal and behavioral representations while explicitly modeling the synergistic relationships between behaviors and modalities, thereby enhancing recommendation performance.
    \item Extensive experiments on benchmark datasets demonstrate that M$^3$BSR significantly outperforms state-of-the-art methods.
\end{itemize}

\section{Related work}
\subsection{Multi-modal Sequential Recommendation}
In recommender systems, multi-modal information is increasingly used to enhance accuracy and capture user preferences. MM-rec~\cite{wu2021mm} extracts image ROIs and employs co-attentional Transformers for text-ROI modeling, alongside a crossmodal attention network for user modeling. UniSRec~\cite{wang2023missrec} leverages item descriptions and contrastive pre-training for universal sequence representations. MISSRec~\cite{wang2023missrec} introduces a Transformer-based encoder-decoder for multi-modal synergy and adaptive item representations. MMSBR \cite{zhang2023beyond} enhances session-based recommendations with pseudo-modality contrastive learning and a hierarchical transformer. MMMLP \cite{liang2023mmmlp} proposes an MLP-based architecture with Feature and Fusion Mixers, achieving state-of-the-art performance with linear complexity. CMCLRec \cite{xu2024cmclrec} addresses user cold-start via cross-modal mapping and simulated behavior sequences. IISAN \cite{fu2024iisan} adopts a Decoupled PEFT structure for intra- and inter-modal adaptation, matching full fine-tuning performance with reduced GPU memory usage.  

\subsection{Multi-behavior Sequential Recommendation}
The utilization of multi-behavior data has emerged as a crucial research direction to enhance recommendation precision and capture users' complex preferences. NMTR \cite{gao2019neural} learns from multi-behavior data by modeling cascading relationships among behaviors and optimizing within a multi-task learning framework. DIPN \cite{guo2019buying} improves real-time purchasing intent prediction by incorporating touch-interactive behavior and using a hierarchical attention mechanism. Feedrec \cite{wu2022feedrec} unifies explicit and implicit feedbacks to infer both positive and negative user interests, enhancing news feed recommendation. SGDL~\cite{gao2022self} introduces a novel denoising paradigm that utilizes memorized interactions during early training to enhance the robustness of recommendation models by guiding subsequent training with adaptive denoising signals. MB-STR \cite{yuan2022multi} models multi-behavior sequential dynamics with a multi-behavior transformer layer and behavior-aware prediction module. KMCLR \cite{xuan2023knowledge} enhances multi-behavior recommendation through knowledge graphs and contrastive learning tasks. EBM \cite{han2024efficient} addresses efficiency and noise in multi-behavior sequential recommendation with hard and soft denoising modules. MISSL \cite{wu2024multi} combines multi-behavior and multi-interest modeling using a hypergraph transformer network and self-supervised learning.  

\section{PRELIMINARY}
\subsection{Problem Statement} Let the user set be $U = \{u_1, u_2,\cdots\}$. Following the work of \cite{bian2021denoising, gao2022self}, we denote clicks as implicit feedback and favorites as explicit feedback. We define the set of behaviors \( B = \{ cl, fa \} \), where \( cl \) represents clicking behavior and \( fa \) represents favoring behavior. Let the item set be $I = \{i_1, i_2,\cdots, i_n\}$.  Each item $i$ has three modalities: item identifier $id$, image $im$, and text $te$. We denote the set of modalities as $M=\{id, im, te\}$. For each user $u$, there are two behavior sequences, namely $S_{u,cl}=(s_{u,cl,1}, s_{u,cl,2},\cdots)$ and $S_{u,fa}=(s_{u,fa,1}, s_{u,fa,2},\cdots)$, where $s_{u,cl,1}$ represents the clicking behavior of user $u$ at time $1$ and $s_{u,fa,1}$ represents the favoring behavior of user $u$ at time $1$.  And for each behavior sequence, since each item has three modalities, there are three corresponding modality sequences.  For example, the item identifier sequence for the clicking behavior of user $u$ can be denoted as $S_{u,cl}^{id}=(s_{u,cl,1}^{id}, s_{u,cl,2}^{id},\cdots)$, the image sequence as $S_{u,cl}^{im}=(s_{u,cl,1}^{im}, s_{u,cl,2}^{im},\cdots)$, and the text sequence as $S_{u,cl}^{te}=(s_{u,cl,1}^{te}, s_{u,cl,2}^{te},\cdots)$. The same applies to the favoring behavior sequences $S_{u,fa}^{id}$, $S_{u,fa}^{im}$, and $S_{u,fa}^{te}$. The core task of this study is to accurately predict the next item $\hat{i}$ that user $u$ will interact with based on the user's multi-modal multi-behavior sequences $S_{u,cl}$ and $S_{u,fa}$. 

\subsection{Diffusion Model} 
\label{sec:conditional diffusion model}
The general conditional diffusion model~\cite{ho2020denoising} consists of a forward diffusion process and a reverse denoising process. The forward diffusion process is a Markov chain. At each time step $t$, according to the variance schedule $\beta_1,\cdots,\beta_T$, Gaussian noise is gradually added to the data $x_0$ to obtain $h_t$, and the process can be expressed as the following formula:  
\begin{equation}  
q(h_t|h_{t-1})=\mathcal{N}(h_t;(1-\beta_t)h_{t-1},\beta_t\mathbf{I})  
\end{equation}  
Where $\mathcal{N}$ denotes the normal distribution, $\beta_t$ represents the variance parameter at time step $t$, and $\mathbf{I}$ is the identity matrix. This equation describes the probability distribution of the current state $h_t$ given the previous state $h_{t-1}$ at time step $t$. This distribution is a normal distribution with a mean of \( (1 - \beta_t)h_{t-1} \) and a covariance of \( \beta_t \mathbf{I} \). This means that $h_t$ is obtained by adding a certain amount of Gaussian noise to $h_{t-1}$, with the amount of noise controlled by $\beta_t$.

The reverse denoising process focuses on learning an inverse transformation to successfully recover the original data $x_0$ from the noisy data $h_t$, and its definition is as follows:  
\begin{equation}
p(h_{t-1} | h_t) = \mathcal{N}(h_{t-1}; \mu_{t-1}(h_t, h^c), \Sigma_{t-1}(h_t, h^c))
\end{equation}
Where $p(h_{t-1} | h_t)$ is the probability distribution of the previous state $h_{t-1}$ given the current state $h_t$. $\mathcal{N}$ denotes the normal distribution. $\mu_{t-1}(h_t, h^c)$ is the predicted mean function based on the current state $h_t$ and conditional information $h^c$. $\Sigma_{t-1}(h_t, h^c)$ is the predicted covariance function based on the current state $h_t$ and conditional information $h^c$.

\begin{figure*}[htbp]
\centering
\includegraphics[width=1\textwidth]{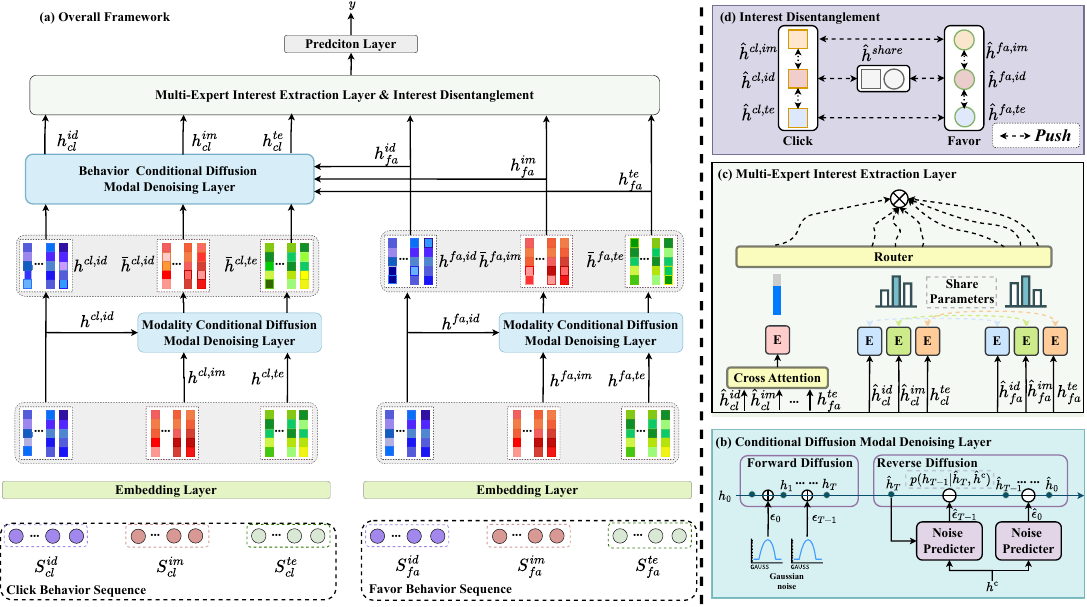}
\caption{(a) The overall framework of our proposed M$^3$BSR framework, which illustrates the complete algorithmic workflow; (b) shows our proposed CDMD Layer; (c) presents our proposed MEIE module; (d) demonstrates the feature decoupling process for different features across various modalities and behaviors.}
\label{fig:framework}

\end{figure*}

\section{Methodology}

In this section, we propose the M$^3$BSR (Multi-Modal Multi-Behavior sequential recommendation) framework to address challenges in characterization of modal preferences across different behaviors and mitigating implicit noise in modalities and behaviors. M$^3$BSR consists of three core components: the Conditional Diffusion Modality Denoising Layer, which enhances multi-modal feature representation by capturing synergistic, common, and specific characteristics; the Conditional Diffusion Behavior Denoising Layer, which improves the signal-to-noise ratio of user behavior representations by leveraging deep-level behavioral information; and the Multi-Expert Interest Extraction Layer, which models common and specific interests across behaviors and modalities to enhance recommendation performance. The framework's structure is illustrated in Fig. \ref{fig:framework}, providing a comprehensive solution for multi-modal multi-behavior sequential recommendation tasks.

\subsection{Multi-modal Multi-behavior Input Layer}  
Multiple different behavior sequences and multiple modalities in each behavior sequence are taken as inputs. For the behavior sequence $S_u$ of user $u$, the three modalities of $id$, image $im$, and text $te$ of each item corresponding to each behavior $s_{u,t}$ need to be processed separately.  
Given the challenges faced by existing multi-behavior recommendations, the information of different modalities has significantly different impacts on different user behaviors. It is crucial to accurately extract and fuse this modal information. First, for the $id$ modality sequence $S^{id}_{u} = [S^{id}_{u,cl}, S^{id}_{u,fa}]$ of user $u$, due to their simplicity, we directly use a embedding layer for embedding:  
\begin{equation}  
h^{id}=MLP(S^{id}_{u})
\end{equation}  
For the image modality sequence $S^{im}_{u} = [S^{im}_{u,cl}, S^{im}_{u,fa}]$ of $u$, we use the Contrastive Language-Image Pre-training (CLIP)~\cite{radford2021learning} to extract features. Let the convolution operation be $CLIP$. After the convolution operation on the image $im_i$, the image feature vector $h^{im}$ is obtained, and the process can be expressed as:  
\begin{equation}  
h^{im} = CLIP(S^{im}_{u})  
\end{equation}  
For the text modality sequence $S^{te}_{u} = [S^{te}_{u,cl}, S^{te}_{u,fa}]$ of $u$, we also use CLIP for embedding text information. After processing by the CLIP, the text feature vector $h^{te}$ is obtained:  
\begin{equation}  
h^{te} = CLIP(S^{te}_{u})
\end{equation}

\subsection{Conditional Diffusion Model Denoising Layer}  

In this section, to address the noise issues in modal and behavioral features, we introduce the Conditional Diffusion Model Denoising (CDMD) Layer for denoising. 

\subsubsection{Denoising for Different Modalities}
As observed in previous works~\cite{wang2023missrec,zhang2023beyond,bian2023multi}, image and text embeddings often contain modality-specific noise that is independent of user preferences. This noise can propagate further through the recommendation pipeline, amplifying the challenges of effectively aligning multi-modal and multi-behavior information. Currently, mainstream recommendation systems are ID-based, as ID features more directly reflect user preferences compared to the complex features of images and texts. Inspired by conditional diffusion model~\cite{ho2020denoising}, we leverage ID features as conditions to guide the denoising of text and image modalities. The specific steps are as follows:  
Let the noisy image and text feature vectors at step $t$ be $\hat{h}^{im}_t$ and $\hat{h}^{te}_t$ respectively. According to the conditional diffusion model in Section~\ref{sec:conditional diffusion model}, the denoising process can be expressed as:  

\textbf{Forward Diffusion:}
\begin{equation}  
q(h_{t+1}^{m}|h_{t}^{m})=\mathcal{N}(\hat{h}_{t+1}^{m};(1-\beta_t)h_{t}^{m},\beta_t\mathbf{I})  
\end{equation} 

Where $m \in \{im, te\}$. $\hat{h}_t^{m}$ is the feature of modality $m$ at time step $t$. $\mathcal{N}$ denotes the normal distribution, $\beta_t$ represents the variance parameter at time step $t$, and $\mathbf{I}$ is the identity matrix. This means that $\hat{h}_t^{m}$ is obtained by adding a certain amount of Gaussian noise to $\hat{h}_{t-1}^{m}$, with the amount of noise controlled by $\beta_t$. By using the reparameterization trick~\cite{ho2020denoising}, the formula can be simplified to:
\begin{equation}
\widetilde{h}_{t+1}^{m} = \sqrt{\alpha_t^{m}}h_{t}^{m} + \sqrt{1-\alpha_t^{m}}\epsilon_{t}^{m}
\end{equation}

Where $\epsilon_{t}^{m}$ represents the noise vector injected into feature $\widetilde{h}_{t+1}^{b,m}$ at time step $t$. $\alpha_t^{m}$ controls the degree of noise added to $m$ at time step $t$. 

\textbf{Reverse Diffusion:}
According to~\cite{ho2020denoising}, the Reverse Diffusion Equation can be expressed as:
\begin{equation}
\begin{aligned}
\label{eq:denosing_modality}
\widebar{h}_{t-1}^{m} &= \frac{1}{\sqrt{\alpha_{t}^{m}}} \left( \widetilde{h}_{t}^{m} - \frac{1 - \alpha_{t}^{m}}{\sqrt{1 - \bar{\alpha}_{t}}} \hat{\epsilon}_{t}^{m} \right)
\end{aligned}
\end{equation}
where $\hat{\epsilon}_{t}^{m}$ represents the noise vector removed from $\widetilde{h}_{t}^{m}$ at time step $t$, and  $\bar{\alpha}_{t} = \prod_{i=1}^{t} \alpha_i$. Inspired by IP-Adapter~\cite{ye2023ip}, we employ conditional features and cross-attention mechanism~\cite{lin2022cat} to guide the removal of noise from the feature $\widetilde{h}_{t}^{m}$. The $\hat{\epsilon}_{t}^{m}$ can be calculated by following equation:
\begin{equation}
\begin{aligned}
\hat{\epsilon}_{t}^{m} &= CI(h_{t}^{m}, h^{c}) \\
CI(\widebar{h}_{t}^m, h_t^c) &= CrossAttention(h_{t}^m, h^{c}, h^{c})
\end{aligned}
\end{equation}

Here, since the ID features can more directly reflect user preferences and contain less noise compared to complex image and text features, we have $h^c = h^{id}$. Note that, for the favor behavior, we denote the features obtained through Reverse Diffusion as $h_{fa}^{im}, h_{fa}^{te}, and h_{fa}^{id}$ for image, text, and ID modalities, respectively.

\textbf{Optimization:} The reconstruction loss is used to measure the difference between the denoised and noisy image and text feature vectors, that is:  
\begin{equation}  
\mathcal{L}_{m} = \sum_{t = 1}^{T} \sum_{m}^{\hat{M}} \left( \| \widetilde{h}_{t}^{m} - \widebar{h}_{t}^{m} \|_2^2 + \| \widetilde{h}_{t}^{m} - \widebar{h}_{t}^{m} \|_2^2 \right)  
\end{equation}  
where $\hat{M} = {im, te}$,$\widetilde{h}_{t}^{m}$ and $\widetilde{h}_{t}^{m}$ are the noisy image and text feature vectors, and $\widebar{h}_{t}^{m}$ and $\widebar{h}_{t}^{m}$ are the denoised feature vectors. By minimizing the $\mathcal{L}_{m}$, it can be ensured that the denoised modal features more accurately reflect users' true preferences. 

\subsubsection{Denoising for Different Behaviors}
Since the noise levels and impacts of different behaviors vary, for example, favor behaviors usually better reflect users' long-term interests and true preferences, while click behavior may contain more noise and temporary factors. click behavior may be caused by users' casual browsing or misoperations and do not fully represent users' real needs, while favor behaviors are decisions made after certain thinking and comparison and are more valuable for reference. Therefore, by denoising click behavior with favor behaviors as conditions, users' true interests can be more accurately captured, and the accuracy and stability of recommendations can be improved. Hence, in the conditional diffusion behavior denoising layer, explicit feedback (such as favor) are used as conditions to denoise implicit feedback behaviors (such as click). The specific steps are as follows:  
Let the feature vector of the implicit feedback behavior be $x_b$, and the denoised behavior feature vector be $x_{b}^{}$. The conditional diffusion behavior denoising process can be expressed as:  

\textbf{Forward Diffusion:}
\begin{equation}
\widetilde{h}_{t+1}^{cl,m} = \sqrt{\alpha_t^{cl,m}}\widebar{h}_{t}^{cl,m} + \sqrt{1-\alpha_t^{cl,m}}\epsilon_{t}^{cl,m}
\end{equation}

Where $\widetilde{h}_{t}^{cl,m}$ is the feature of behavior click for modality $m$ at time step $t$. $\widebar{h}_{t}^{cl,m}$ is the denoising feature by Eq.~(\ref{eq:denosing_modality}) for modality $m$ of behavior click. $\alpha_t^{cl, m}$ controls the degree of noise added to behavior click for $m$ at time step $t$. 

\textbf{Reverse Diffusion:}
\begin{equation}
\begin{aligned}
\hat{h}_{t-1}^{cl,m} &= \frac{1}{\sqrt{\alpha_{t}^{cl,m}}} \left( \widetilde{h}_{t+1}^{cl,m}- \frac{1 - \alpha_{t}^{cl,m}}{\sqrt{1 - \bar{\alpha}_{t}^{cl,m}}} \hat{\epsilon}_{t}^{cl,m} \right)\\
\hat{\epsilon}_{t}^{cl,m} &= CI(h_{t}^{cl,m}, h^{c}) \\
CI(h_{t}^{cl,m}, h_t^c) &= CrossAttention(h_{t}^{cl,m}, h^{c}, h^{c})
\end{aligned}
\end{equation}

Compared to click behavior, which may be influenced by accidental actions or other factors, deeper behaviors (such as favoring) tend to exhibit a lower level of noise. Therefore, we can use the features of favor behaviors to guide the denoising of click behavior, we have $h^c = h^{fa,m}$. Note that, for the click behavior, we denote the features obtained through Reverse Diffusion as $h_{cl}^{im}, h_{cl}^{te}, and h_{cl}^{id}$ for image, text, and ID modalities, respectively.

\textbf{Optimization:} Similarly, the reconstruction loss is used to measure the difference between the denoised and noisy behavior feature vectors, that is:  
\begin{equation}  
\mathcal{L}_{b} = \sum_{m}^{\hat{M}}\sum_{t = 1}^{T} \| \hat{h}_{t}^{cl,m} - \widetilde{h}_{t}^{cl,m} \|_2^2  
\end{equation}  
where $\widetilde{h}_{t}^{cl,m}$ is the noisy behavior feature vector, and $\hat{h}_{t}^{cl,m}$ is the denoised feature vector for modality $m$ of behavior click. By minimizing this loss, the model can remove the noise in the click behavior features, more accurately model users' real behavior, and thus improve the accuracy of recommendations.

\subsection{Multi-Expert Interest Extraction Layer}

In this section, we introduce the Multi-Expert Interest Extraction (MEIE) Layer, a novel approach designed to capture and aggregate diverse user interests from multi-modal and multi-behavior data. This layer effectively handles the complexity of user interactions by leveraging cross-attention mechanisms, Transformer architectures, and gating networks.

\subsubsection{Common Feature Extraction}
Users exhibit diverse behaviors and interaction modalities, yet often have a stable underlying preference. For example, a fan of science fiction movies keeps this preference whether browsing text-based movie info or watching video trailers. To capture user preferences, we must extract common features across different behaviors and modalities. Since these features are initially independent, we use a cross - attention mechanism. It dynamically calculates feature correlations, captures semantic relationships, and effectively aligns and fuses heterogeneous feature representations.

Specifically, the multi-modal features of click and favor behaviors are concatenated firstly:
\begin{equation}
h_{cl} = [h_{cl}^{im}, h_{cl}^{te}, h_{cl}^{id}]
\end{equation}
\begin{equation}
h_{fa} = [h_{fa}^{im}, h_{fa}^{te}, h_{fa}^{id}]
\end{equation}
where $h_{cl/fa}^{im/te/id}$ denotes the features obtained through the \text{reverse diffusion} process.

Then, cross-attention is used to interact and align between click and favor behaviors:
\begin{equation}
\hat{h}^{share} = \text{CrossAttention}(h_{cl}, h_{fa})
\end{equation}

Finally, considering that the Transformer can effectively capture long-range dependencies and complex interactions, we utilize the Transformer as the expert network to capture deeper common features:
\begin{equation}
h_{common} = \text{Transformer}(\hat{h}^{share})
\end{equation}
where $h_{com}$ represents the common features across different behaviors and modalities.

\subsubsection{Unique Feature Extraction for Click and Favor Behavior}
In practical applications, apart from common features, users may also be attracted by the characteristics of modalities on products. For instance, users might focus more on the visual attractiveness of images or the accuracy of text descriptions. Moreover, user click behavior and favor behavior may exhibit distinct preferences across different modalities. Hence, extracting the unique modal features under different behaviors is essential for a better understanding of users' specific interests for different modalities. Given the Transformer's capability to dynamically attend to diverse sequence positions and capture intricate intra-modal dependencies, it effectively learns behavior-specific information representations. Thus, we employ the Transformer as the expert network to extract distinct modal features (image, text, and ID) for click and favor behaviors.

\begin{equation}
h_{cl}^{m} = \text{Transformer}(h_{cl}^{m})
\end{equation}
\begin{equation}
h_{fa}^{m} = \text{Transformer}(h_{fa}^{m})
\end{equation}

Here, $h_{cl}^{m}$, and $h_{fa}^{m}$ represent the modality $m$ features under click and favor behavior, respectively. $m \in \{id, im, te\}$.

\subsubsection{Interest Disentanglement:} To ensure that the feature representations of different modalities and behaviors can be distinguished and capture their unique characteristics, we design a contrastive loss function for feature disentanglement. This encourages the feature representations $h_{cl}^{id}$, $h_{cl}^{im}$, $h_{cl}^{te}$, $h_{fa}^{id}$, $h_{fa}^{im}$, $h_{fa}^{te}$, and $h_{common}$ to be separated in the feature space. The purpose of this design is to avoid confusion between features and ensure that each modality and behavior's feature representation can independently capture its unique semantic information. Specifically, We use the following contrastive loss function to encourage the separation of different feature representations:
\begin{equation}
\mathcal{L}_{\text{contrast}} = -\sum_{i \neq j} \log \frac{\exp(\text{sim}(h_i, h_j) / \tau)}{\sum_{k \neq i} \exp(\text{sim}(h_i, h_k) / \tau)}
\end{equation}

where \( h_i \) and \( h_j \) represent different feature representations (e.g., \( h_{cl}^{id} \), \( h_{cl}^{im} \), \( h_{cl}^{te} \), \( h_{fa}^{id} \), \( h_{fa}^{im} \), \( h_{fa}^{te} \), and \( h_{common} \)). \(\text{sim}(h_i, h_j)\) is a similarity measure between features \( h_i \) and \( h_j \), typically using cosine similarity. \(\tau\) is a hyperparameter, known as the temperature parameter, which controls the sensitivity to similarity in contrastive learning.

\subsubsection{Interest Routing Fusion}

Users may exhibit varying preferences across different modalities. Additionally, the feature data often contains noise unrelated to user interests, which can interfere with the accurate extraction and modeling of true user preferences. Therefore, we employ a routing network to dynamically control the weights of different features during fusion. This network prioritizes features that are more aligned with user preferences and less affected by noise, ensuring a more robust and precise representation of user interests. The routing network can be represented as:
\begin{equation}
\begin{aligned}
    h_{cl} &= [h_{cl}^{im}, h_{cl}^{te}, h_{cl}^{id}] \\
    h_{fa} &= [h_{fa}^{im}, h_{fa}^{te}, h_{fa}^{id}] \\
    g = \sigma(W_g \cdot &[h_{common}; h_{cl}; h_{fa}] + b_g)
\end{aligned}
\end{equation}
\begin{equation}
y = g \cdot h_{common} + (1 - g) \cdot [h_{cl}; h_{fa}]
\end{equation}
here, $\sigma$ is the sigmoid function, $W_g$ and $b_g$ are learnable parameters, and $y$ is the final fused feature. 

\subsection{Prediction \& Optimization}

\subsubsection{cross-entropy loss:} After completing the multi-modal multi-behavior interest extraction, the user preference representation $y$ is input into the prediction layer. This layer uses a fully-connected neural network to map the preference representation to the probability distribution of items:

\begin{equation}
p(i|y)=\text{softmax}(W y + b)
\end{equation}

To train the model, we use the cross-entropy loss function to measure the difference between the predicted results and the actual results. The specific formula is as follows:

\begin{equation}
\mathcal{L}_{\text{main}}=-\sum_{u\in U}\log p(i_{u}|y_{u})
\end{equation}

where $i_{u}$ is the actual item selected by user $u$, and $y_{u}$ is the user preference representation extracted by the model for user $u$.

\subsubsection{Total Loss Function:} 
The total loss function is:  
\begin{equation}  
\label{eq:loss_all}
\mathcal{L} = \mathcal{L}_{main} + \lambda_c \mathcal{L}_{\text{contrast}} + \lambda_m \mathcal{L}_{m} + \lambda_b \mathcal{L}_{b}  
\end{equation}  
where $\lambda_c$, $\lambda_m$, and $\lambda_b$ are hyperparameters that balance the contributions of the contrastive loss, modal denoising loss, and behavior denoising loss, respectively. Minimizing the total loss function  ensures that the model comprehensively captures user preferences while reducing noise and enhancing feature representations. This holistic approach leads to more robust and accurate recommendations.

\section{Experiment}
In this section, we evaluate the M$^3$BSR framework on two public datasets, Rec-Tmal and Kuaishou, to address the following research questions:
\begin{itemize}
\item \textbf{Effectiveness (RQ1).} Does the M$^3$BSR model outperform various state-of-the-art (SOTA) baselines?
\item \textbf{Thoroughness (RQ2).} How do the specific designs in M$^3$BSR impact the model's performance?
\item \textbf{Robustness (RQ3).} How do changes in the module's parameters affect the model's effectiveness?
\item \textbf{Cold Start (RQ4).} How is the performance of our method under the cold start scenario?
\item \textbf{Visualization (RQ5).} Can our method effectively remove noise?
\end{itemize}

\begin{table}[]
\caption{Statistics of Rec-Tmal and Kuaishou datasets.}

\setlength{\tabcolsep}{1mm}{
\begin{tabular}{c|c|c|c|c|c}
\hline
\textbf{Dataset} & \textbf{Users} & \textbf{Items} & \textbf{Interactions} & \textbf{Avg. Clicks} & \textbf{Avg. Favors} \\ \hline
Rec-Tmal         & 72051          & 93466        & 328387             & 4.16               & 1.73               \\
Kuaishou         & 22793         & 618529     & 5852725          & 255.02            & 40.96             \\ \hline
\end{tabular}}
\label{tab:datasets}

\end{table}

\subsection{Experimental Settings}

\begin{table*}[]
\caption{Performance Comparison on Rec-Tmal and Kuaishou datasets.}

\setlength{\tabcolsep}{2mm}{
\begin{tabular}{c|cccc|cccc}
\hline
\multirow{2}{*}{\textbf{Method}} & \multicolumn{4}{c|}{\textbf{Rec-Tmal}} & \multicolumn{4}{c}{\textbf{Kuaishou}} \\ \cline{2-9}
& \textbf{HR@10} & \textbf{NDCG@10} & \textbf{HR@20} & \textbf{NDCG@20} & \textbf{HR@10} & \textbf{NDCG@10} & \textbf{HR@20} & \textbf{NDCG@20} \\ \hline
\multicolumn{9}{c}{\textbf{Traditional Sequential Recommendation Methods}} \\ \hline
GRU4Rec (2015) & 0.1498 & 0.0431 & 0.2015 & 0.0492 & 0.0853 & 0.0021 & 0.1360 & 0.0103 \\
SASRec (2018) & 0.1662 & 0.0578 & 0.2145 & 0.0635 & 0.1001 & 0.0083 & 0.1490 & 0.0141 \\
BERT4Rec (2019) & 0.2123 & 0.0964 & 0.2629 & 0.1078 & 0.1435 & 0.0494 & 0.1958 & 0.0549 \\
STOSA (2022) & 0.2568 & 0.1370 & 0.3080 & 0.1420 & 0.1930 & 0.0908 & 0.2420 & 0.1135 \\
SSDRec (2024) & 0.2827 & 0.1637 & 0.3315 & 0.1680 & 0.2176 & 0.1151 & 0.2671 & 0.1198 \\ \hline
\multicolumn{9}{c}{\textbf{Multi-modal Sequential Recommendation Methods}} \\ \hline
MISSRec (2023) & 0.2786 & 0.1601 & 0.3322 & 0.1623 & 0.2179 & 0.1120 & 0.2654 & 0.1186 \\
MMSBR (2023) & 0.2897 & 0.1660 & 0.3336 & 0.1702 & 0.2209 & 0.1188 & 0.2711 & 0.1253 \\
MMMLP (2023) & 0.2928 & 0.1753 & 0.3421 & 0.1789 & 0.2284 & 0.1285 & 0.2769 & 0.1332 \\
M3SRec (2023) & 0.2997 & 0.1803 & 0.3486 & 0.1851 & 0.2315 & 0.1293 & 0.2859 & 0.1354 \\
CMCLRec (2024) & 0.3075 & \underline{0.1886} & \underline{0.3602} & 0.1937 & 0.2394 & 0.1375 & \underline{0.2918} & \underline{0.1450} \\ \hline
\multicolumn{9}{c}{\textbf{Multi-behavior Sequential Recommendation Methods}} \\ \hline
NMTR (2019) & 0.2738 & 0.1524 & 0.3225 & 0.1563 & 0.2096 & 0.1061 & 0.2583 & 0.1088 \\
DIPN (2019) & 0.2786 & 0.1602 & 0.3307 & 0.1654 & 0.2153 & 0.1111 & 0.2623 & 0.1156 \\
SGDL (2022) & 0.2861 & 0.1670 & 0.3359 & 0.1708 & 0.2212 & 0.1181 & 0.2672 & 0.1209 \\
MB-STR (2022) & 0.2931 & 0.1741 & 0.3447 & 0.1795 & 0.2290 & 0.1282 & 0.2794 & 0.1344 \\
KMCLR (2023) & 0.3025 & 0.1800 & 0.3518 & 0.1857 & 0.2372 & 0.1334 & 0.2806 & 0.1382 \\
EBM (2024) & \underline{0.3083} & 0.1877 & 0.3601 & \underline{0.1953} & \underline{0.2420} & \underline{0.1395} & 0.2882 & 0.1423 \\ \hline
\textbf{M$^3$BSR (Ours)} & \textbf{0.3207} & \textbf{0.2028} & \textbf{0.3815} & \textbf{0.2130} & \textbf{0.2603} & \textbf{0.1630} & \textbf{0.3101} & \textbf{0.1612} \\ \hline
\end{tabular}}
\label{tab:AUC_compared_SOTA_new}

\end{table*}

\subsubsection{Dataset.} Due to the scarcity of research specifically focused on multi - modal multi - behavior recommendation systems, we introduced two new datasets for our experiments: "Rec-Tmal" and "Kuaishou". The Rec-Tmal dataset\footnote{\url{https://tianchi.aliyun.com/dataset/140281}}  is sourced from Tmall\footnote{\url{https://www.tmall.com/}},  and the Kuaishou dataset\footnote{\url{https://www.kuaishou.com/activity/uimc/datadesc}}  comes from the Kuaishou platform. These two datasets offer diverse multi-modal and multi-behavior data, which are essential for comprehensively evaluating the performance of our proposed M$^3$BSR model in such a novel research area.  For the two datasets, we utilized product images for visual information representation and product titles for textual information. For Rec-Tmal, the ID information included item ID, brand ID, user ID, and seller ID. Each user in this dataset has two types of behaviors: click and favor. For Kuaishou, the ID modality includes user ID and video ID. The detailed parameters are shown in Table~\ref{tab:datasets}

For all datasets, we selected user behavior sequences with a minimum length of 5. Additionally, we retained the 50 most recent historical records for each user. For the training and test data, we adopt the same setting as described in~\cite{zhou2018deep, xiao2022abstract}. Table~\ref{tab:datasets} displays the relevant statistical information for each dataset. Following~\cite{yang2022multi, yuan2022multi, han2024efficient}, we define the target behavior as "favor" and use "click" as the auxiliary behavior, enabling the model to capture the relationship between different user behaviors and their impact on recommendations. To ensure a realistic evaluation scenario, we treat the last favor behavior in each user sequence as the test sample and the preceding ones as validation samples, allowing the model to predict future user preferences based on historical interactions. Additionally, to enhance the robustness of our evaluation, we pair each positive sample with 99 randomly selected negative instances, as suggested in~\cite{yang2022multi, yuan2022multi, han2024efficient}.

\subsubsection{Evaluation Metrics.}
In our evaluation, we employ Hit Rate at $K$ (HR@$K$) and Normalized Discounted Cumulative Gain at $K$ (NDCG@$K$) to assess prediction quality, widely accepted in sequential recommendation. HR@$K$ measures whether the target item appears in the top-$K$ recommendations, while NDCG@$K$ evaluates the item's ranking position, prioritizing higher ranks. We use $K=10$ and $20$ to comprehensively test the model's performance across varying recommendation lengths.

\subsubsection{Implementation Details.} The proposed model is implemented using the PyTorch framework\footnote{https://pytorch.org}. To ensure a fair comparison, we utilize our pipeline framework to reproduce all of the baselines, and each baseline model is experimented with multiple times to obtain optimal results. The size of IDs modality is set to 16, and image and text modalities embeddings are set to 512 for the complexity of image and text features compared to IDs. We use a fixed mini-batch size of 1024. When searching for optimal values, we explore learning rates in the set \{10$^{-5}$, 10$^{-4}$, 10$^{-3}$\}, and hidden sizes in the set \{64, 128, 256, 512\}. The time step $T$ for the diffusion models is set to 15. The level of the diffusion noise, denoted as $\alpha$, are initialized in the range of [0.001, 0.1] to ensure effective noise scheduling during training. Additionally, we search for the weights in Eq.~(\ref{eq:loss_all}) within the range of 0.01 to 0.15. To prevent overfitting and optimize performance, we employ an early stopping strategy. Specifically, if the NDCG@10 metric does not improve for 10 consecutive epochs, the training process will be halted.

\subsubsection{Comparison Methods.} We compare our framework with SOTA sequential recommendation methods across three categories. Traditional Sequential Recommendation Methods: GRU4Rec~\cite{hidasi2015session}, SASRec~\cite{kang2018self}, BERT4Rec~\cite{sun2019bert4rec}, STOSA~\cite{fan2022sequential}, SSDRec~\cite{zhang2024ssdrec}; Multi-modal Sequential Recommendation Methods: MISSRec~\cite{wang2023missrec}, MMSBR~\cite{zhang2023beyond}, MMMLP~\cite{liang2023mmmlp}, M3SRec~\cite{bian2023multi}, CMCLRec~\cite{xu2024cmclrec}; Multi-behavior Sequential Recommendation Methods: NMTR~\cite{gao2019neural}, DIPN~\cite{guo2019buying}, SGDL~\cite{gao2022self}, MB-STR~\cite{yuan2022multi}, KMCLR~\cite{xuan2023knowledge}, EBM~\cite{han2024efficient}.

\subsection{Performance Comparison (RQ1)}

To validate the effectiveness of our proposed model, we performed experiments on two different datasets. The results, presented in Table~\ref{tab:AUC_compared_SOTA_new}, compare the performance of M$^3$BSR with that of the baseline models across different categories. Based on these results, we made the following observations:

\begin{itemize}
\item \textbf{M$^3$BSR consistently achieves the best performance in terms of HR@10, HR@20, NDCG@10, and NDCG@20 across both datasets.} This demonstrates the superior effectiveness of our proposed framework in handling multi-modal multi-behavior sequential recommendation tasks. Besides, the consistent gains across different datasets further underscore the robustness and generalizability of M$^3$BSR.

\item \textbf{Compared to traditional sequential recommendation methods (GRU4Rec, SASRec, BERT4Rec, STOSA, SSDRec), M$^3$BSR shows significant improvements.} This is attributed to M$^3$BSR's ability to leverage multi-modal information and model inter-behavior dependencies, which are not considered by these simpler sequential models.

\item \textbf{M$^3$BSR also outperforms state-of-the-art multi-modal sequential recommendation methods (MISSRec, MMSBR, MMMLP, M3SRec, CMCLRec).} The improvements highlight the effectiveness of M$^3$BSR's conditional diffusion denoising mechanisms in removing noise from both modal and behavioral representations, leading to more accurate user preference modeling. For example, while methods like MISSRec focus on multi-modal synergy, they may not explicitly address noise in the same way as M$^3$BSR.

\item \textbf{Furthermore, M$^3$BSR surpasses multi-behavior sequential recommendation methods (NMTR, DIPN, SGDL, MB-STR, KMCLR, EBM).} This indicates the advantage of M$^3$BSR in jointly modeling multi-modalities and multi-behaviors. While methods like NMTR focus on cascading behavior relationships, they lack the explicit handling of multi-modal information and noise that M$^3$BSR provides.
\end{itemize}

\begin{figure}[htbp]
\centering
\includegraphics[width=0.45\textwidth]{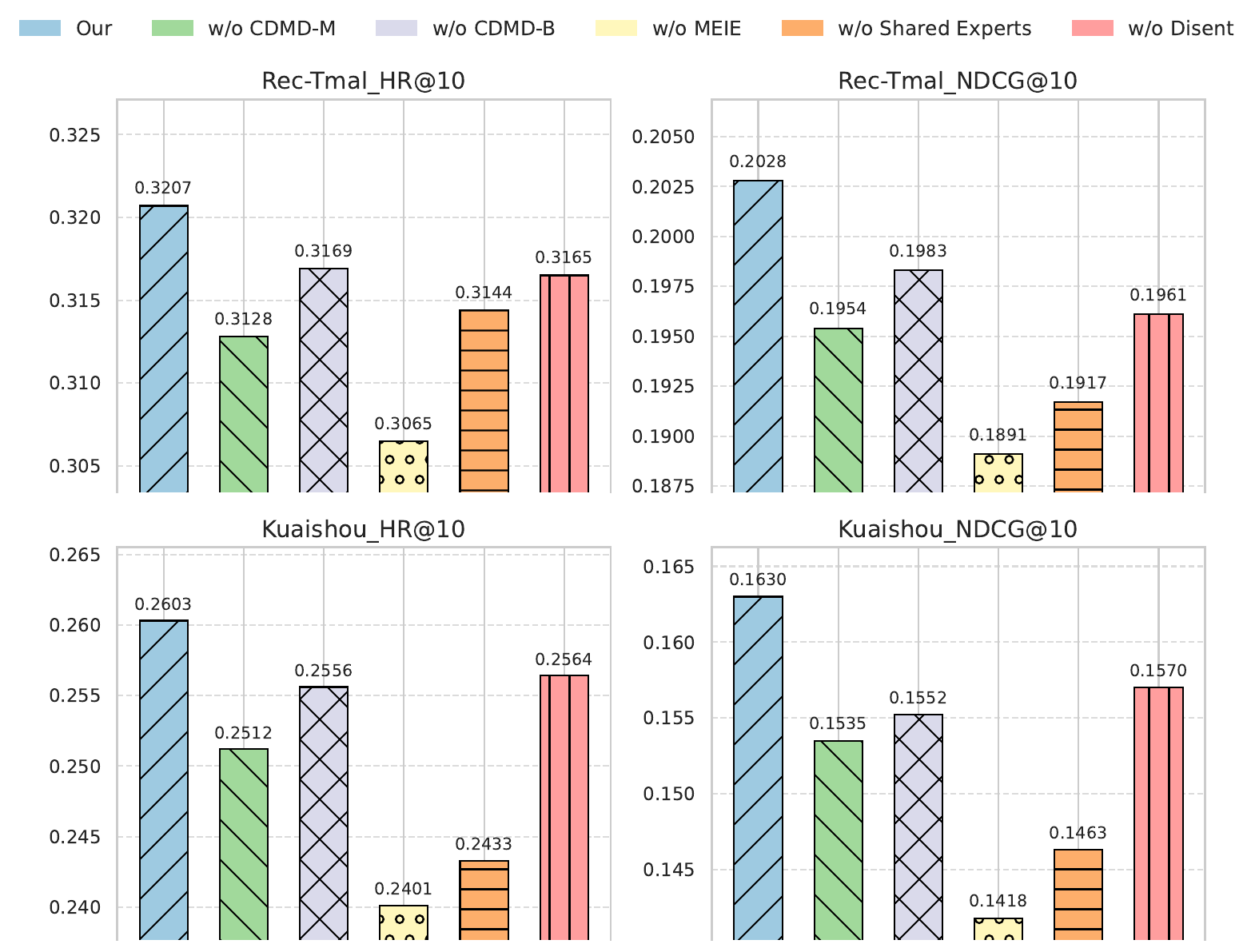}

\caption{Performance for Ablation Study}
\vspace{-0.5cm}
\label{fig:Ablation Study}
\end{figure}

\subsection{Ablation Study (RQ2)}
In this section, we conduct an ablation study to evaluate the impact of different components of the M$^3$BSR framework on the performance of recommender systems.

\begin{itemize}
\item \textbf{w/o CDMD-M}: We remove the Conditional Diffusion Model Denoising (CDMD) module for modalities, which is responsible for denoising multi-modal feature representations.
\item \textbf{w/o CDMD-B}: We remove the CDMD module for behaviors, which denoises behavior feature representations.
\item \textbf{w/o MEIE}: We remove the Multi-Expert Interest Extraction (MEIE) layer that extracts and decouples user interests from multi-modal and multi-behavior data.
\item \textbf{w/o Shared Expert}: We remove the shared expert network, which plays a role in modeling common interests across modalities and behaviors.
\item \textbf{w/o Disent}: We remove the Interest Disentanglement module from the proposed model.
\end{itemize}

The ablation study in Fig.~\ref{fig:Ablation Study} evaluates the impact of key modules on recommendation performance. (1) Removing the CDMD for module modalities reduces HR@10 and NDCG@10, demonstrating the importance of denoising multi-modal features for accurate user preference modeling. (2) Similarly, eliminating the CDMD module for behaviors degrades performance, emphasizing the need to mitigate noise in behavior sequences. (3) The most significant performance drop occurs when removing the MEIE Layer, highlighting its critical role in modeling common and specific interests across modalities and behaviors. (4) Ablating the Shared Expert also reduces performance, as it disrupts the modeling of synergistic relationships between data sources. (5) Finally, removing the Interest Disentanglement module diminishes performance, confirming its role in separating feature representations and reducing confusion across modalities and behaviors.

\subsection{In-depth Analysis (RQ3)}

\begin{figure}[htbp]
    \centering
    \begin{subfigure}[t]{0.15\textwidth}
        \centering
        \includegraphics[width=1\textwidth]{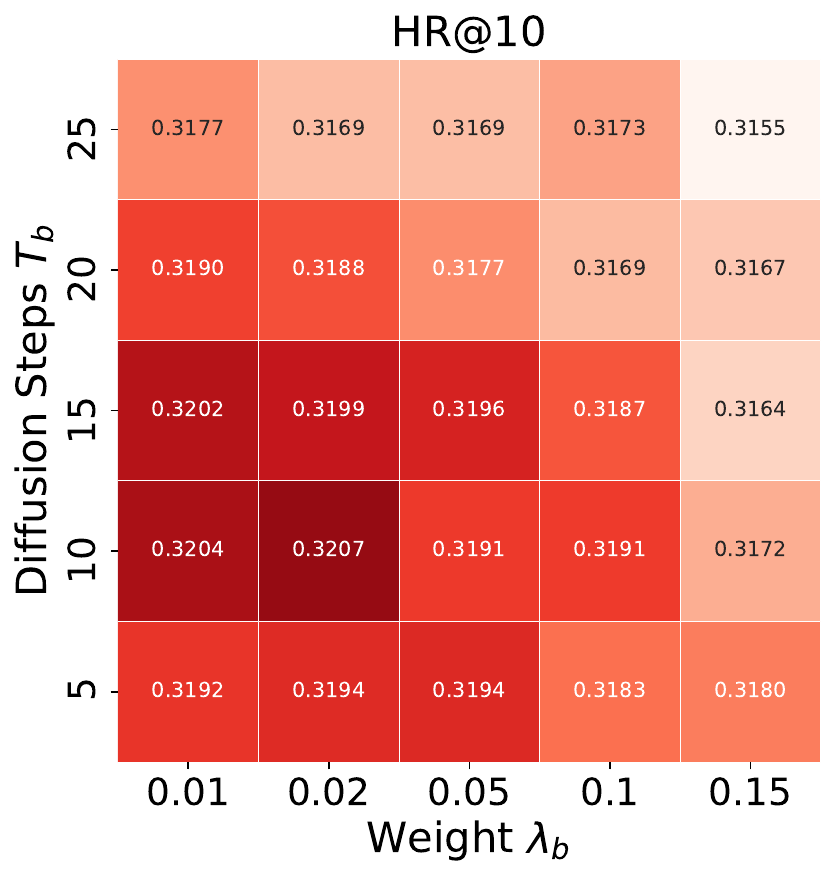}
        \caption{Rec-Tmal:$T_b$ \& $\lambda_b$}
        \label{fig:heatmap1_taobao}
    \end{subfigure}
    \begin{subfigure}[t]{0.15\textwidth}
        \centering
        \includegraphics[width=1\textwidth]{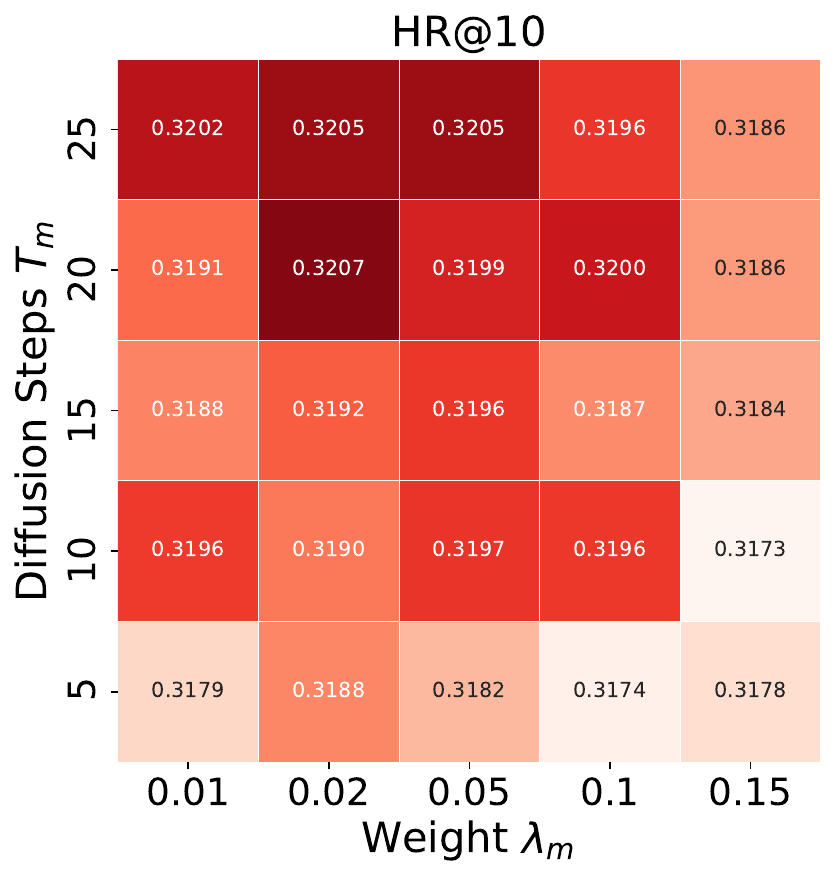}
        \caption{Rec-Tmal:$T_m$ \& $\lambda_m$}
        \label{fig:heatmap2_taobao}
    \end{subfigure}
    \begin{subfigure}[t]{0.15\textwidth}
        \centering
        \includegraphics[width=1\textwidth]{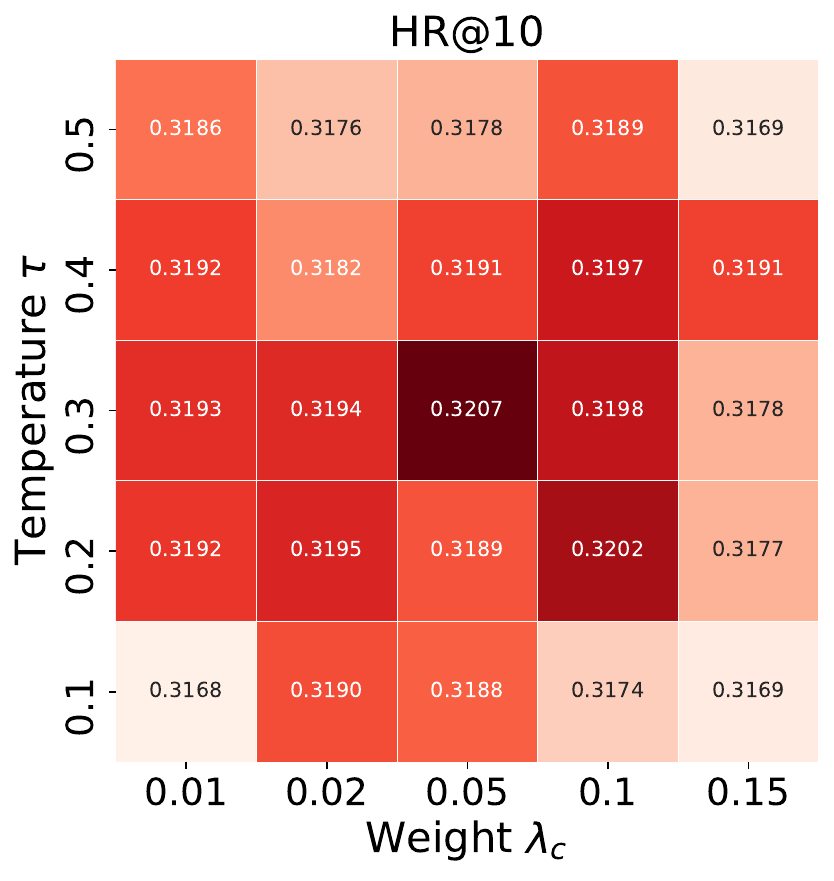}
        \caption{Rec-Tmal:$\tau$ \& $\lambda_c$}
        \label{fig:heatmap3_taobao}
    \end{subfigure}
    \begin{subfigure}[t]{0.15\textwidth}
        \centering
        \includegraphics[width=1\textwidth]{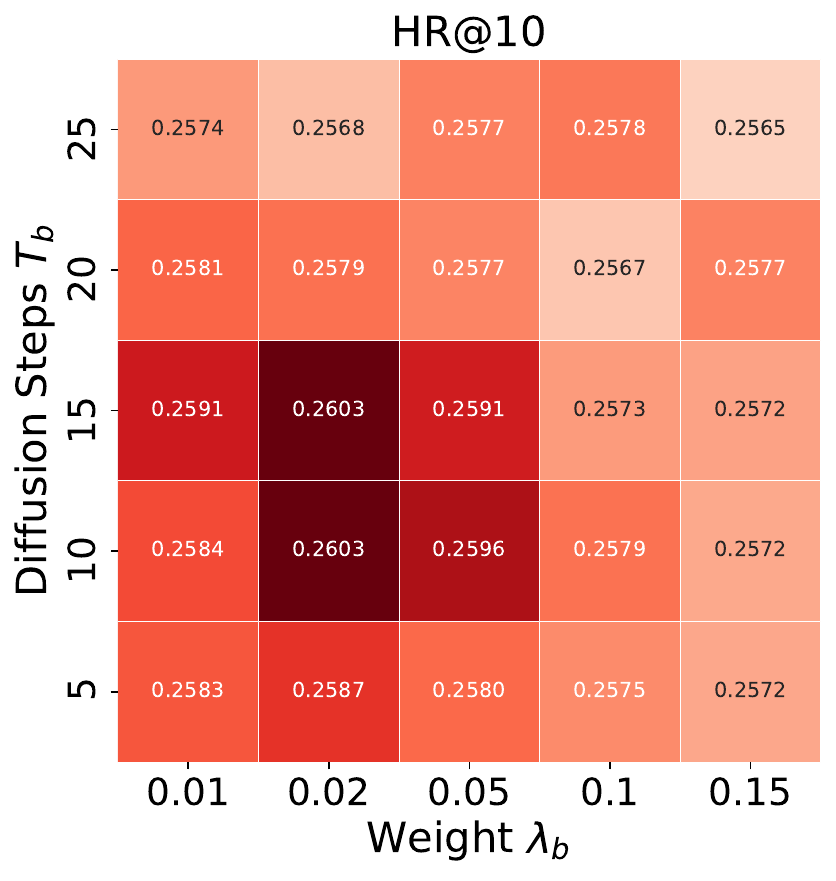}
        \caption{Kuaishou:$T_b$ \& $\lambda_b$}
        \label{fig:heatmap1_kuaishou}
    \end{subfigure}
    \begin{subfigure}[t]{0.15\textwidth}
        \centering
        \includegraphics[width=1\textwidth]{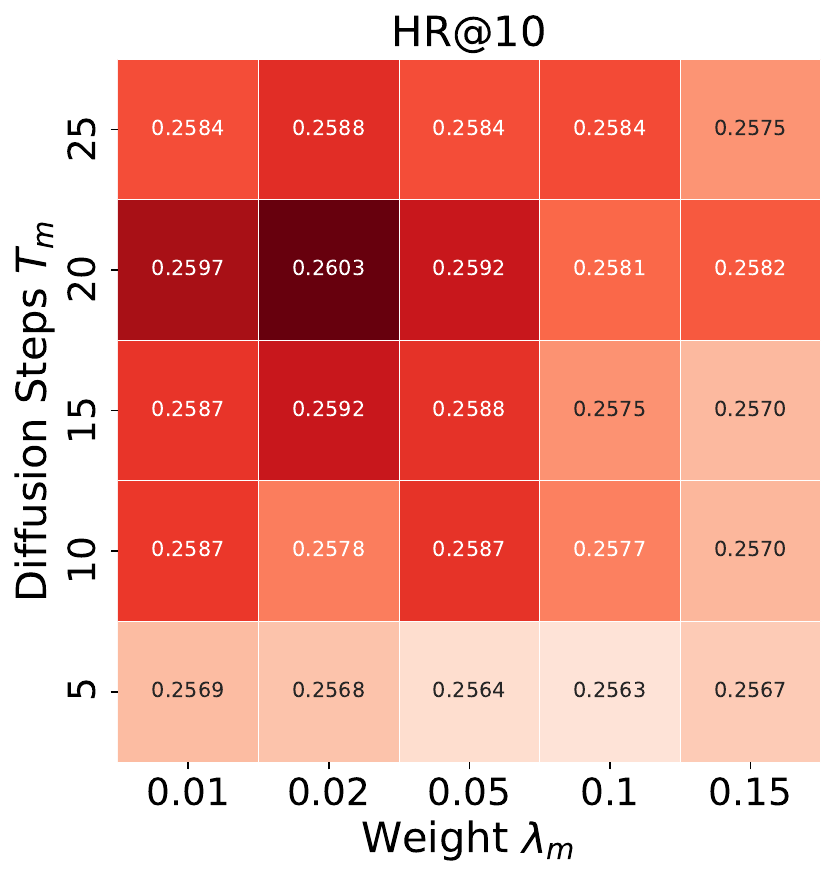}
        \caption{Kuaishou:$T_m$ \& $\lambda_m$}
        \label{fig:heatmap2_kuaishou}
    \end{subfigure}
    \begin{subfigure}[t]{0.15\textwidth}
        \centering
        \includegraphics[width=1\textwidth]{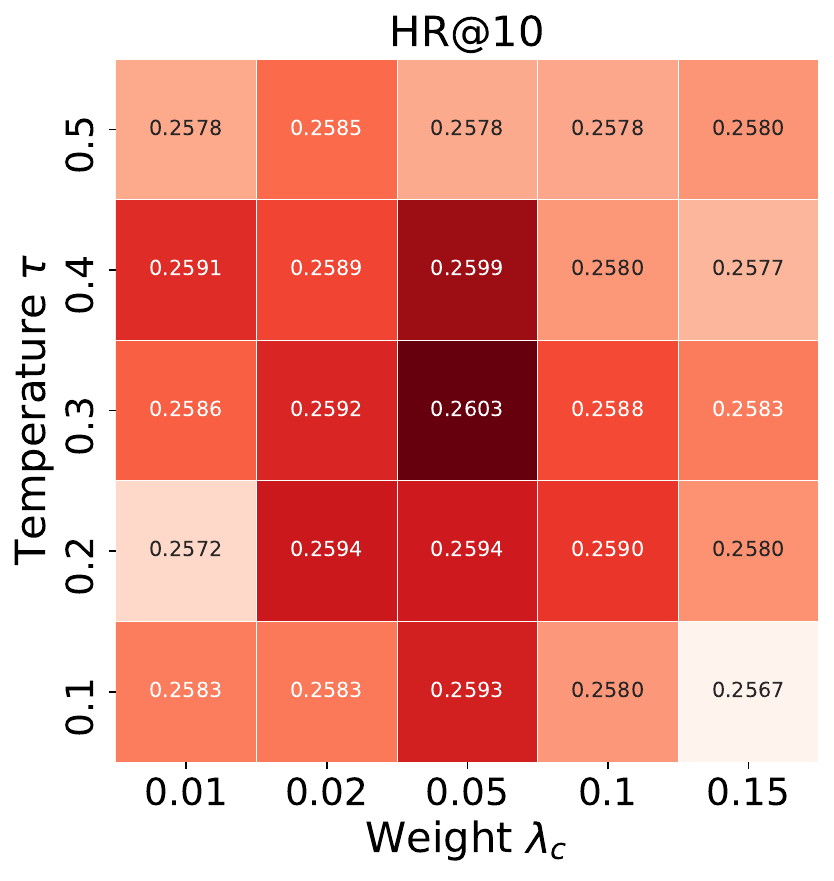}
        \caption{Kuaishou:$\tau$ \& $\lambda_c$}
        \label{fig:heatmap3_kuaishou}
    \end{subfigure}
    \vspace{-0.2cm}
    \caption{Heatmaps for different weights and diffusion steps}
    \vspace{-0.5cm}
    \label{fig:heatmaps}
\end{figure}

\subsubsection{Effect of Behavior Denoising Diffusion}
Figs.\ref{fig:heatmap1_taobao} and\ref{fig:heatmap1_kuaishou} show HR@10 performance for varying behavior denoising steps and loss weights on the Rec-Tmal and Kuaishou datasets. On Rec-Tmal, the peak HR@10 (0.3207) occurred at 10 steps / 0.02 weight, balancing denoising and specificity; excessive steps/weight (e.g., 25 / 0.15) caused over-regularization (HR@10 0.3155). Similarly, Kuaishou achieved its best HR@10 (0.2603) at 10 steps / 0.02 weight, ensuring effective denoising without over-regularization. Insufficient denoising (e.g., 5 steps / 0.01) reduced performance (HR@10 0.2583). Consistently, the optimal range across both datasets is around 10 diffusion steps with a 0.02 loss weight, achieving the best trade-off.

\subsubsection{Effect of Modality Denoising Diffusion}
Figs.\ref{fig:heatmap1_taobao} and\ref{fig:heatmap1_kuaishou} show HR@10 performance for varying behavior denoising steps and loss weights on the Rec-Tmal and Kuaishou datasets. On Rec-Tmal, the peak HR@10 (0.3207) occurred at 10 steps / 0.02 weight, balancing denoising and specificity; excessive steps/weight (e.g., 25 / 0.15) caused over-regularization (HR@10 0.3155). Similarly, Kuaishou achieved its best HR@10 (0.2603) at 10 steps / 0.02 weight, ensuring effective denoising without over-regularization. Insufficient denoising (e.g., 5 steps / 0.01) reduced performance (HR@10 0.2583). Consistently, the optimal range across both datasets is around 10 diffusion steps with a 0.02 loss weight, achieving the best trade-off.

\subsubsection{Effect of Contrastive Learning}
Figs.\ref{fig:heatmap3_taobao} and\ref{fig:heatmap3_kuaishou} show the interplay between temperature and contrastive loss weight. On Rec-Tmal, peak HR@10 (0.3207) occurred at 0.3 temperature / 0.05 weight, balancing sample separation and alignment. Lower settings (e.g., 0.01 temp/0.01 weight) caused incorrect pairings (HR@10 0.3168), while higher settings (e.g., 0.5 temp/0.15 weight) hindered separation (HR@10 0.3169). Similarly, Kuaishou's best HR@10 (0.2603) was at 0.3 temp / 0.05 weight, ensuring proper separation. Lower settings (0.01/0.01) dropped HR@10 to 0.2583, and higher ones (0.5/0.15) to 0.2580. Consistently, optimal performance across both datasets requires a 0.3 temperature and 0.05 loss weight.

\subsection{Cold Start Experiment (RQ4)}
In cold-start evaluations on Kuaishou users with sparse history ($\leq$10 interactions), M$^3$BSR demonstrated significantly better performance (HR@10, NDCG@10) than strong baselines. This advantage arises from its ability to effectively infer preferences from multi-modal data (images/text via CDMD), identify broader modality-based interests independent of deep interaction history (using MEIE \& Interest Disentanglement), and reduce noise impact through feature denoising.

\begin{table}[]
\caption{Performance on Cold-Start Users (Rec-Tmal)}
\vspace{-0.2cm}
\setlength{\tabcolsep}{3mm}{
\begin{tabular}{c|cc}
\hline
\textbf{Method} & \textbf{HR@10} & \textbf{NDCG@10} \\ \hline
STOSA          & 0.2186         & 0.1459           \\
MMMLP           & 0.2413         & 0.1672           \\
EBM             & 0.2560         & 0.1819           \\
\textbf{M$^3$BSR (Ours)} & \textbf{0.2897} & \textbf{0.2146}  \\ \hline
\end{tabular}}
\label{tab:cold_start}

\end{table}

\subsection{TSNE Visualization (RQ5)}

\begin{figure}[htbp]
\centering
\includegraphics[width=0.45\textwidth]{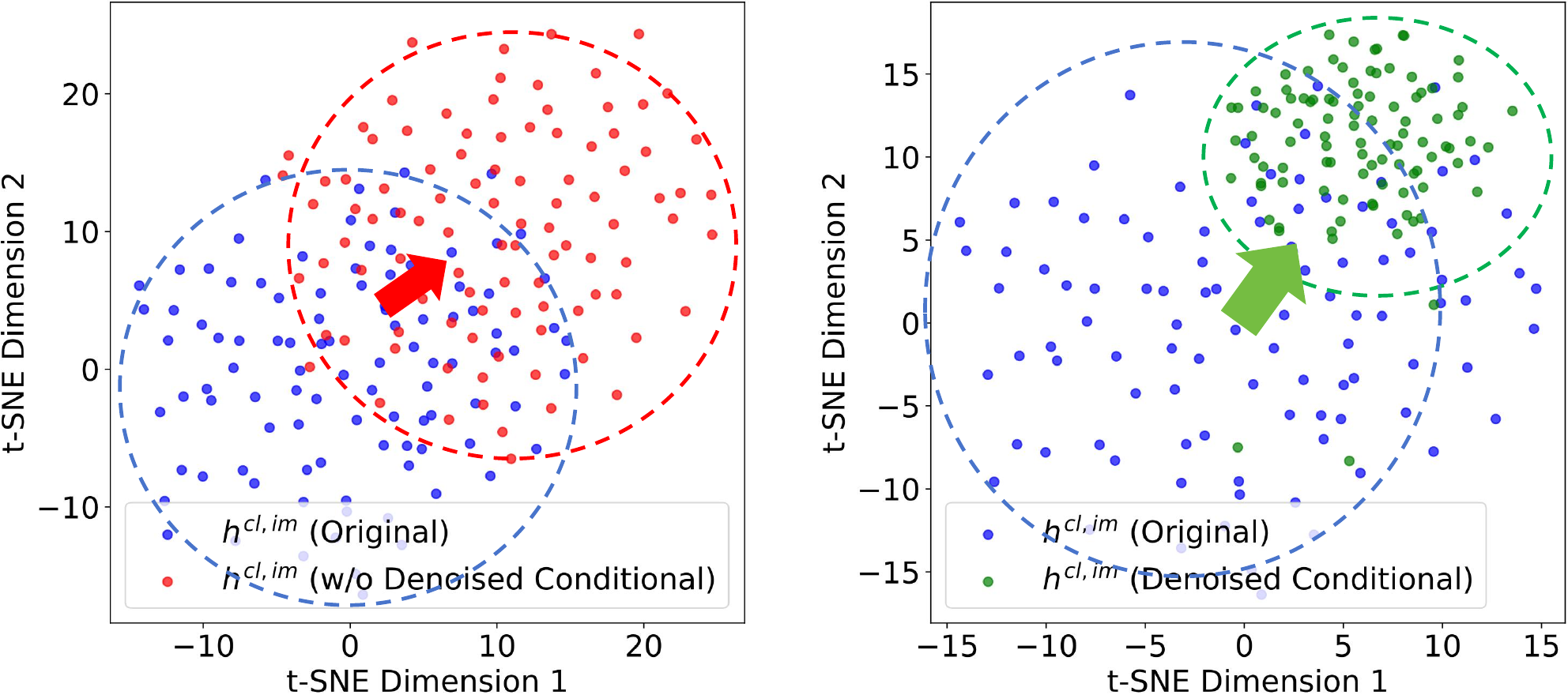}

\caption{Visualization of CMDM Effectiveness}
\label{fig:tsne}
\end{figure}

To visualize the effectiveness of the CDMD module, we performed t-distributed stochastic neighbor embedding (t-SNE) on the item embeddings learned from the Kuaishou dataset. As shown in Fig.~\ref{fig:tsne}, the embeddings processed by CDMD exhibit compact clusters, indicating that CDMD effectively reduces noise. In contrast, when CDMD is replaced with a Multilayer Perceptron (MLP), the embeddings only undergo translational shifts without significant changes in their overall distribution. The resulting clusters remain dispersed and noisy, failing to achieve the denoising effect observed with CDMD.

\subsection{Time Efficiency Experiment}
M$^3$BSR's time efficiency was evaluated on Kuaishou. We fix the batch-size as 128. Table~\ref{tab:time_efficiency} shows the average training time per epoch and the average inference time per prediction for each method.While its added complexity increases training time, this is manageable given significant performance gains. Its inference time is comparable to complex baselines, making it practical for online recommendation due to its parallelizable modular design.

\begin{table}[]
\caption{Time Efficiency Comparison on Kuaishou Dataset}
\vspace{-0.2cm}
\setlength{\tabcolsep}{2mm}{
\begin{tabular}{c|c|c}
\hline
\multirow{2}{*}{\textbf{Method}} & \textbf{Training Time} & \textbf{Inference Time} \\ 
                & \textbf{per Epoch (s)} & \textbf{per Prediction (s)} \\ \hline
STOSA & 25.3 & 0.07 \\
MMMLP & 28.7 & 0.09 \\
EBM & 32.5 & 0.13 \\
\textbf{M$^3$BSR (Ours)} & 35.2 & 0.16 \\ \hline
\end{tabular}}
\label{tab:time_efficiency}

\end{table}

\section{Conclusion}
This paper proposes M$^3$BSR to address multi-modal recommendation challenges. Using specialized modules (CDMD for Modalities/Behaviors, MEIE \& Feature Decoupling) and contrastive loss, it denoises inputs and models common/specific interests. These components work together to denoise multi-modal features, mitigate noise in user behavior sequences, and explicitly model common and specific interests across modalities and behaviors. Experiments show M$^3$BSR outperforms baselines, improving preference modeling and recommendation accuracy.
\newpage
\bibliographystyle{ACM-Reference-Format}
\balance
\bibliography{sample-base}

\appendix

\end{document}